\documentclass[conference]{IEEEtran}
\usepackage[utf8]{inputenc}
\IEEEoverridecommandlockouts

\usepackage{cite}
\usepackage{amsmath,amssymb,amsfonts}
\usepackage{algorithmic}
\usepackage{graphicx}
\usepackage{textcomp}
\usepackage{xcolor}
\def\BibTeX{{\rm B\kern-.05em{\sc i\kern-.025em b}\kern-.08em
    T\kern-.1667em\lower.7ex\hbox{E}\kern-.125emX}}
    
\begin{document}

\title{Positioning Fog Computing for \\ Smart Manufacturing
}

\author{
    \IEEEauthorblockN{
        1\textsuperscript{st} Jaakko Harjuhahto}
        \IEEEauthorblockA{\textit{Department of Computer Science} \\
        \textit{Aalto University}\\
        Espoo, Finland\\
        jaakko.harjuhahto@aalto.fi
    }
    \and
    \IEEEauthorblockN{
        2\textsuperscript{nd} Vesa Hirvisalo}
        \IEEEauthorblockA{\textit{Department of Computer Science} \\
        \textit{Aalto University}\\
        Espoo, Finland\\
        vesa.hirvisalo@aalto.fi
    }
}

\maketitle

\begin{abstract}


We study machine learning systems for real-time industrial quality control. In many factory systems, production processes must be continuously controlled in order to maintain product quality. Especially challenging are the systems that must balance in real-time between stringent resource consumption constraints and the risk of defective end-product. There is a need for automated quality control systems as human control is tedious and error-prone. We see machine learning as a viable choice for developing automated quality control systems, but integrating such system with existing factory automation remains a challenge. In this paper we propose introducing a new fog computing layer to the standard hierarchy of automation control to meet the needs of machine learning driven quality control.


\end{abstract}


\section{Introduction}
\label{sec_introduction}

Environmental sustainability and resource-awareness are central issues for the modern global world. For the manufacturing industries, digitalization \cite{chen2020} has the potential to positively contribute by increasing resource and information efficiency as a result of applying emerging IIoT (Industrial Internet of Things) technologies.  But, negative impacts can also be foreseen as digitalization causes shifts in resource and energy use as well as waste and emissions, and encourages rapid disposal of obsolete systems.

There is also significant inertia in the manufacturing industries themselves. The current factory systems are very complex, deeply integrated and already heavily optimized, and re-engineering the systems will require massive investments into the technology itself. Viewed from the reality of a factory floor \cite{jbrair2018}, applying modern IIoT technologies such as machine learning into quality control of production has plenty of challenges \cite{abdelzaher2020}.

In this paper, we discuss computing solutions for using machine learning in the industrial production systems, where the production processes must be controlled continuously in order to keep up the quality of the product. 

We review the problems faced in direct application of fog technology in factory setups. The current computing solutions for factories do not offer the scalable high-performance real-time computing with rich software support that is needed for modern approaches based on machine learning. Also, direct application of generic fog technologies is hard as they lack true integration into factories and do not comply with the related non-functional requirements (e.g., real-time).

We propose an architectural solution that is based on having a fog computing layer merged into traditional factory systems. We identify three dimensions of essential architectural decisions in designing middleware for our architectural solution: node resource management, fog layer management, and virtualization. Considering these three dimensions, we discuss our solution in the context of existing literature.



The structure of this paper is the following. We start with an overview of the technologies that are needed for the resource-aware digitalization of manufacturing (Section \ref{sec_background}). We continue in Section \ref{sec_problem_statement} by reviewing the problems faced, when modern machine learning technologies are applied in factories for quality control. After that, we describe shortly our previous research and concentrate on describing and discussing our architectural solution (Section \ref{sec_approach}). We end our presentation with our short conclusions.

\section{Background}
\label{sec_background}

Our work connects to several areas of ongoing research. Factory systems have a long tradition, to which digitalization \cite{zhong2017} has entered through a long evolution of technology. Recently, novel artificial intelligence technologies have emerged on the scene especially in the form of deep learning. This has caused urgent need to increase both the richness of software structures and the available computing power at industrial factory sites and the supporting infrastructure. In this section, we briefly overview factory systems, deep learning, real-time systems, and generic fog computing solutions.

Factories have computing solutions that differ significantly from the systems of typical consumer devices and the related supporting infrastructure. Factory systems are Cyber-Physical Systems (CPS) that include both digital and physical parts. They often use PLC (Programmable Logic Controller) technology for automation. Classical PLC is hardware-based and needs supporting systems to operate. For factories, it is typical to use SCADA (Supervisory Control and Data Acquisition) that controls the manufacturing processes and does monitoring, acquisition and processing real-time data. \cite{jbrair2018}

Deep learning (DL) is form of machine learning that uses neural networks with multiple layers to progressively extract higher-level features from raw input \cite{murshed2022}. From the application development viewpoint, using deep learning technologies includes massive computation for training the networks, and then, separate inference by using the trained networks. Thinking of factory setups, the inference computing should be done close to the raw data sources (e.g., sensors). 
Inference often calls for high performance computation with accelerators, which can exhibit complex computational behavior \cite{MMSys-accel}. Further, complex inference approaches \cite{zhuoqing2021} are needed, when there are multiple heavy sensor streams with real-time processing requirements as is typical for large factories.

Real-time requirements add significant complications for the design of computing solutions. Real-time systems \cite{davis2011} are often classified into soft real-time systems, for which timing requirements are not strict, and hard real-time systems, for which correct timing behavior is a must. From the software viewpoint, this calls for precise understanding of the timing behavior of the software on top of a specific hardware platform \cite{wilhelm2008}. Having multiple and/or heterogeneous processing systems, which is typical for deep learning, adds significant complexity to the computing solutions \cite{davis2011}. Recently, using online learning for IIoT systems has gained attention \cite{shahhosseini2022}.

The fog computing paradigm addresses the ways of bridging the gap between devices and centralized cloud services, utilizing resources in between them, and thus allowing sufficient resources close to the devices. In addition to cloud computing, fog computing is closely connected to the edge computing paradigm. In edge computing, the data is processed in the control of the network edge (e.g., base station) and typically on the sensor device itself without being transferred anywhere. For fog and edge computing, it is essential to understand how to organize for learning and inference, data processing, quality assurances, distribution of computation (e.g., offloading), and organizing the control \cite{abdelzaher2020}.

%

\section{Problem Statement}
\label{sec_problem_statement}

We view machine learning and big data as major motivations for adopting a fog computing paradigm for smart manufacturing applications. Both are computationally expensive methods to integrate into a cyber-physical environment, where timing behavior matters. The requirements of deep learning applications are beyond the performance capabilities of PLCs, which are the backbone of time critical computing for automation \cite{jbrair2018}. Using cloud-based infrastructure is unfeasible because of data volumes, latency requirements, and reliability concerns.

A common ad-hoc solution is to deploy individual industrial PCs to service these compute needs. With significant resource over-provisioning, they can offer sufficient timing behavior but this approach scales poorly, both in terms of capacity and management, as applications with DL components proliferate. Dedicated embedded systems offer the best-in-class timing behavior, but embedded development suffers from even worse scaling challenges. Adopting to evolving requirements is also needed, as data science is inherently exploratory and iterating over designs is common, since better data sets are compiled over time. This scaling problem can be addressed using cloud native tools, but these solutions are throughput oriented and not intended to guarantee timing behavior for individual requests, which severely limits their usability for CPS systems. \cite{merlino2019}

The pre-existing installation base of factory automation equipment is enormous, and any solution for deploying DL enhanced applications must synergize with existing technologies. But machine learning enhanced applications, such as machine-vision based quality inspection, are a poor fit for the two standard layers of automation hierarchy, described in Section \ref{sec_background}. The direct control layer has far too limited compute resources to run complex algorithms, and the supervisory control layer that operates on a soft real-time best-effort basis is unsuitable for integration with the CPS control loop. A new type of solution is needed.

\section{Our Approach}
\label{sec_approach}

We introduce a new fog computing layer to the standard hierarchy of automation control. Fog nodes are placed physically and logically close to the demand for compute capabilities, and all the fog platform resources are abstracted and offered as a service. For a smart factory, several micro-datacenters are co-located with the automation equipment. In addition to solving the compute resource issues of deploying machine learning enhanced CPS applications at scale, a fog computing layer offers other benefits, such as fault tolerance through redundancy from the shared pool of resources, and the flexibility to customize and reconfigure the system with only software. 

\begin{figure}
    \begin{center}
        \includegraphics[width=0.8\linewidth]{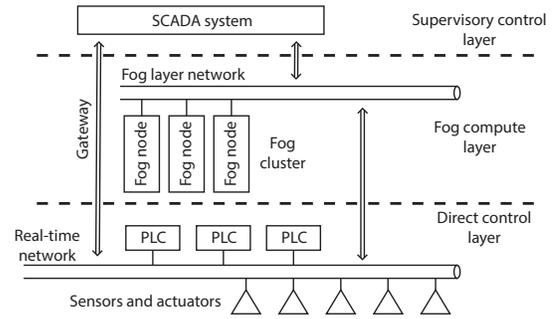}
        \caption{The industrial automation hierarchy. At the top is the supervisory control layer and at the bottom the direct control layer. In between is our proposed fog computing layer.}
        \label{fig_layer_hierarchy}
    \end{center}
\end{figure}

Adding a layer of more sophisticated compute capabilities between the direct control and supervisory control layers can retain a level of real-time behavior but provide sufficiently rich capabilities to run arbitrary complex workloads. Crucially, the fog layer can be added to existing automation systems to provide new capabilities without the need to change previously installed equipment. Figure \ref{fig_layer_hierarchy} illustrates this arrangement.

As the fog computing layer is presented to tenant applications as an abstract pool of resources, the middleware running on individual fog nodes and managing the system as a whole is crucial to the system architecture. Such solutions are pervasive in cloud computing, but existing cloud native solutions are poorly suited for the smart factory domain, as they do not provide a mechanism for ensuring the quality of service of individual tasks and service requests, but view attained quality of service as an amortized metric. In the cyber-physical space, individual tasks and requests matter. Existing cloud native tools retroactively measure service levels, when proactive resource aware methods are needed to ensure predictable timing behavior. 

Our previous work with quality inspection for the manufacturing industry \cite{INDIN-Topo} demonstrated how deep learning based solutions can assist with inspection tasks that have previously proven difficult, e.g., pattern recognition based problems. Our inspection device solved the quality control problem, but the integrated and significant hardware-software stack would have to be replicated for each device within factory premises. Without a common compute infrastructure, the same applies to other DL enhanced applications, leading to an explosion of near identical hardware-software stacks.

A central research question is to identify the best model for offering predictable fog computing as a service for IIoT. Existing models of offering infrastructure, platforms, functions, containers, software, etc as a service \emph{do not currently treat time as a first class concept}. Augmenting existing solutions with proactive resource aware planning remains an option.

For evaluating future solutions, we identify three dimensions of architectural decisions crucial in designing middleware for a CPS enabled fog. The dimensions define a design space to explore for research into predictable and reliable fog computing for industrial use. In the following, we discuss the three dimensions.

\subsection{D-1: Node Resource Management}

For hard real-time embedded systems, a key development phase is the analysis of the application to ascertain its resource utilization. The phase is increasingly difficult as the software stack grows in complexity, as is the case with complex third party domain specific software such as with deep learning libraries. Without a full prescriptive plan available, how to best model resources and perform management to achieve deterministic timing remains a research question. Investigating how much a-priori information and planning is needed to assist the platform in achieving predictable execution is a key challenge.

As complete analysis is often unfeasible, methods based on instrumentation and learning offer a potential approach to implement self-stabilizing middleware. The research in \cite{shahhosseini2022} demonstrates how online learning can adopt a holistically optimal execution strategy for DL tasks.

\subsection{D-2: Fog Layer Management}

Making decisions about where and when on a shared platform to place services and tasks is known as orchestration \cite{costa2022}. For example, Kubernetes \cite{kubernetes} employs a top-down model, where nodes are controlled by a centralized entity that makes all the decisions on orchestration based on policies and constraints defined by developers. This centralized model is suitable for warehouse scale computing, as solving the global packing problem allows operators to maximize hardware utilization, as density is a key criterium for cost effectiveness. Existing research has studied adopting cloud-native IaaS solutions for varied fog use cases \cite{santos2019} and adopting fog for specific industrial applications \cite{li2018}, but does not adequately address integration with real-time CPS systems.

For predictable fog computing, service quality is paramount over density. As each node has control over their local scheduling and other resources on an OS or hypervisor level, the nodes themselves are also best suited to evaluate how they can meet the service requirements for a certain task or service. Integrating scheduling information from individual nodes with cluster level decision making requires a multi-tier solution. Holistic management must also include networking to guarantee deterministic data stream delivery, with Time Sensitive Networking (TSN) \cite{pop2018} offering a way to implement this for packet data. 

\subsection{D-3: Virtualization}

A single commercial off-the-shelf (COTS) fog node can run as either a bare metal host or virtualize the hardware resources by using a hypervisor. Research into real-time middleware for COTS systems, e.g. \cite{rouxel2021}, often builds on top of bare metal host with a modified Linux kernel for real-time scheduling \cite{reghenzani2020}. However, bare metal deployment of software is increasingly being replaced by virtualization to solve isolation and software engineering challenges, e.g., dependency management. Containers present a lightweight solution by virtualizing the host operating system, which can still enforce scheduling \cite{abeni2019} of the containerized application. 

For an industrially viable platform, applications do not necessarily need access to a fully featured desktop kernel, e.g., the Linux kernel. With a focus on offering a platform for predictable execution of computationally expensive work, research can identify opportunities to simplify layers of abstractions to better fit the domain while offering improved facilities to monitor and control resources. An example of this principle is Firecracker \cite{agache2020} implementing a custom virtualization solution specialized for function-as-a-service compute. Striking a balance between maintaining support for arbitrary application code while incorporating lessons learned from dedicated real-time operating systems is a key challenge. 


\section{Conclusions}

In this paper, we discussed challenges of integrating machine learning based applications with existing manufacturing automation systems. We argue that the current standard hierarchy of automation is well suited for existing needs, but is not sufficient for supporting the compute and time behavior requirements of emerging IIoT solutions.

Our proposal is motivated by our own previous research and experience with factory systems.The central idea in our proposal is to add a fog computing layer to the standard hierarchy of automation control to support machine learning based solutions for industrial sensing and control. 


\bibliography{references, urls}
\bibliographystyle{unsrt}  

\end{document}